\documentclass[journal=jpcafh,manuscript=article]{achemso}

\usepackage{booktabs}
\usepackage{etoolbox}
\usepackage[version=4]{mhchem}

\author{Jiachen Li}
\affiliation{Department of Chemistry, Duke University, Durham, NC 27708, USA}
\alsoaffiliation{Department of Chemistry, Yale University, New Haven, CT, USA 06520}
\author{Weitao Yang}
\affiliation{Department of Chemistry, Duke University, Durham, NC 27708, USA}
\email{weitao.yang@duke.edu}

\title{Chemical Potentials and the One-Electron Hamiltonian of the Second-Order Perturbation Theory from the Functional Derivative Approach}

\begin{document}

\begin{tocentry}
\includegraphics[width=1\textwidth]{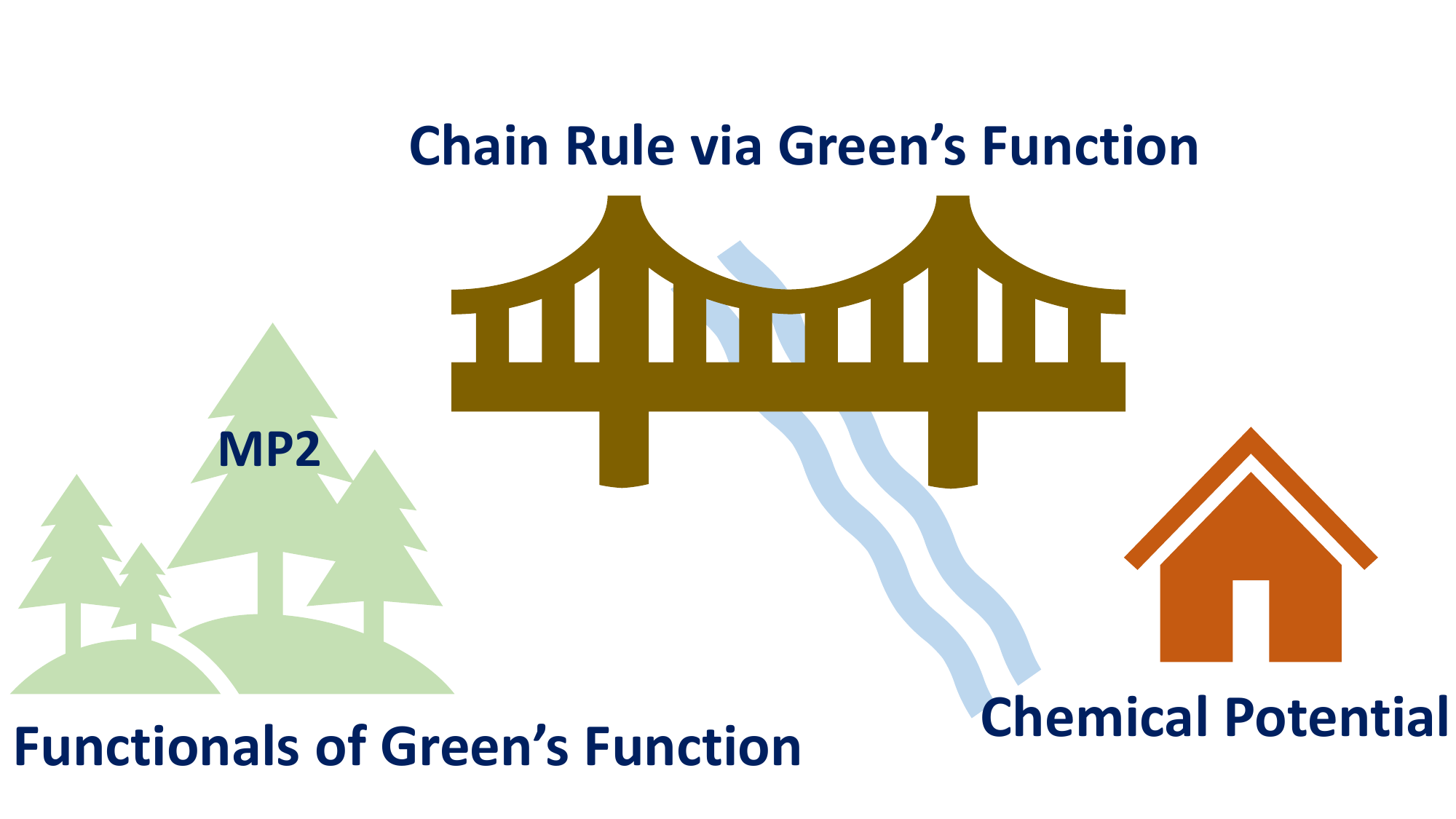}
\end{tocentry}

\begin{abstract}
We develop a functional derivative approach to calculate the chemical potentials of the second-order perturbation theory (MP2). 
In the functional derivative approach,
the correlation part of the MP2 chemical potential,
which is the derivative of the MP2 correlation energy with respect to the occupation number of frontier orbitals,
is obtained from the chain rule via the non-interacting Green's function.
First, the MP2 correlation energy is expressed in terms of the non-interacting Green's function and its functional derivative to the non-interacting Green's function is the second-order self-energy.
Then the derivative of the non-interacting Green's function to the occupation number is obtained by including the orbital relaxation effect.
We show that the MP2 chemical potentials obtained from the functional derivative approach agrees with that obtained from the finite difference approach. 
The one-electron Hamiltonian, 
defined as the derivative of the MP2 energy with respect to the one particle density matrix, 
is also derived using the functional derivative approach, 
which can be used in the self-consistent calculations of MP2 and double-hybrid density functionals.
The developed functional derivative approach is promising for calculating the chemical potentials and the one-electron Hamiltonian of approximate functionals and many-body perturbation approaches dependent explicitly on the non-interacting Green's function.
\end{abstract}

\section{Introduction}
The chemical potential,
defined as the derivative of total energy $E$ with respect to $N$, 
the total number of electrons at the fixed external potential, 
is one of the most important concepts for studying fundamental electronic properties in molecules and materials,
such as the electron transfer and the chemical reactivity\cite{parrDensityFunctionalTheoryAtoms1989}.
The chemical potentials of the electron removal and addition process are equal to the negative of the ionization potential (IP) and the electron affinity (EA),
based on the exact $E(N)$ linear condition\cite{perdewDensityFunctionalTheoryFractional1982,yangDegenerateGroundStates2000}.
Although these quantities can be measured by photoemission and inverse photoemission spectroscopy,
the computational study can provide insights for understanding electronic structures from basic principles.
In past decades,
much efforts have been devoted to develop quantum chemistry approaches to predict the chemical potentials.
Kohn-Sham density functional theory\cite{kohnSelfConsistentEquationsIncluding1965,parrDensityFunctionalTheoryAtoms1989} (KS-DFT),
as the most popular approach in modern quantum chemistry, 
has been widely used for molecular and periodic systems\cite{zhangComparisonDFTMethods2007,chongInterpretationKohnSham2002,zhanIonizationPotentialElectron2003}.
In the (generalized) KS-DFT formalism,
orbital energies of the highest occupied molecular orbital (HOMO) and the lowest unoccupied molecular orbital (LUMO) energy have been shown to be rigorously the chemical potentials for electron removal and electron addition for exchange-correlation energy functionals that are continuous in the KS density matrix\cite{cohenFractionalChargePerspective2008}.
In such cases, 
which include all commonly used exchange-correlation energy functional approximations such as GGA, meta GGA, hybrid, and perturbation-theory based approximation, the HOMO and LUMO energies are well justified to approximate the negative of the IP and the EA, because the chemical potentials of the exact functional are the negative of the IP and the EA based on the linearity condition for fractional electron numbers\cite{perdewDensityFunctionalTheoryFractional1982,cohenFractionalChargePerspective2008}.
For the exact functional, 
it has been shown to be discontinuous for strongly correlated systems\cite{mori-sanchezDiscontinuousNatureExchangeCorrelation2009}.
Therefore, for strongly correlated systems, 
the chemical potentials are equal to the HOMO and LUMO orbital energies plus the discontinuity contributions.
However, 
DFT has an undesired dependence on the density functional approximations (DFAs) and its accuracy is affected by the intrinsic delocalization error\cite{cohenInsightsCurrentLimitations2008,mori-sanchezLocalizationDelocalizationErrors2008}.
Alternatively,
the Green's function formalisms such as $GW$\cite{hedinNewMethodCalculating1965,martinInteractingElectrons2016,reiningGWApproximationContent2018,golzeGWCompendiumPractical2019} and T-matrix\cite{zhangAccurateQuasiparticleSpectra2017,liRenormalizedSinglesGreen2021,orlandoThreeChannelsManybody2023} approximation can be applied for reliable predictions of the chemical potential.
In the Green's function formalism,
the IP and the EA are predicted by the quasiparticle energy that directly measures the charged excitation energy. 
It has been shown that Green's function approaches substantially improve the accuracy of predicting energy levels over the KS-DFT approach for both occupied and unoccupied states,
which are the key quantities to calculate IPs, EAs and core-level binding energies\cite{vansettenGW100BenchmarkingG0W02015,carusoBenchmarkGWApproaches2016,zhangAccurateQuasiparticleSpectra2017,vlcekStochasticGWCalculations2017,vlcekSwiftGW102018,vansettenAssessingGWApproaches2018,wilhelmGWCalculationsThousands2018,loosGreenFunctionsSelfConsistency2018,loosGreenFunctionsSelfConsistency2018,jinRenormalizedSinglesGreen2019,liRenormalizedSinglesGreen2021,zhuAllElectronGaussianBasedG0W02021,liBenchmarkGWMethods2022,liRenormalizedSinglesCorrelation2022,liLinearScalingCalculations2023,panades-barruetaAcceleratingCoreLevelGW2023,venturellaMachineLearningManyBody2024,tolleAcceleratingAnalyticContinuationGW2024}.

The second-order perturbation theory\cite{mollerNoteApproximationTreatment1934,szaboModernQuantumChemistry2012} (MP2) has been a long standing method in the quantum chemistry community.
MP2 has been widely employed for predicting many energetic and geometric properties including the electron density, the geometry and the energy barrier for both molecular and periodic systems\cite{klopperMP2LimitCorrection1999,frankPNOMP2Gradient2017,haserEquilibriumGeometryC601991,feyereisenUseApproximateIntegrals1993,neuhauserExpeditiousStochasticApproach2013}.
The success of MP2 stems from the improvement over the Hartree-Fock (HF) method and the favorable computational cost compared with coupled cluster methods.
In addition,
the MP2 density matrix is broadly used to generate natural orbitals for the quantum embedding theory\cite{nusspickelSystematicImprovabilityQuantum2022,schaferLocalEmbeddingCoupled2021}.
However, 
it is well-known that MP2 fails to describe open-shell systems and transition-metal complexes\cite{grimmeSpincomponentscaledElectronCorrelation2012,soydasAccurateOpenShellNoncovalent2013,neeseAssessmentOrbitalOptimizedSpinComponent2009}.
To address this issue,
various methods based on conventional MP2 including spin-component-scaled MP2\cite{grimmeSpincomponentscaledElectronCorrelation2012,lochanOrbitaloptimizedOppositespinScaled2007,bozkayaOrbitaloptimizedThirdorderMollerPlesset2011} (SCS-MP2), 
orbital optimized MP2\cite{pulayOrbitalinvariantFormulationSecondorder1986,bozkayaAnalyticEnergyGradients2013,neeseAssessmentOrbitalOptimizedSpinComponent2009,bozkayaQuadraticallyConvergentAlgorithm2011} (OO-MP2)
and regularized MP2\cite{sheeRegularizedSecondOrderMoller2021,rettigRevisitingOrbitalEnergyDependent2022} have been developed.
The form of the MP2 correlation energy has also been applied in DFT as an approximation to the G\"{o}rling-Levy second-order perturbation correlation energy\cite{gorlingCorrelationenergyFunctionalIts1993}, 
which leads to the double-hybrid functional.
In the double-hybrid functional calculations,
in addition to the hybridization of the HF exchange, 
the MP2 correlation energy evaluated with KS orbitals is mixed as a certain portion of the correlation energy\cite{goerigkDoublehybridDensityFunctionals2014,sancho-garciaDoublehybridDensityFunctionals2013,zhangDoublyHybridDensity2011}.
A large number of double-hybrid functionals have been developed in past decades,
which significantly outperform conventional DFAs for predicting a broad range of properties including geometries, dissociation energies, thermochemistry and thermochemical kinetics\cite{savinCorrelationContributionsDensity1991,schwabeChemicalAccuracyThermodynamics2006,tarnopolskyDoubleHybridFunctionalsThermochemical2008,benighausSemiempiricalDoubleHybridDensity2008,kartonHighlyAccurateFirstPrinciples2008,chaiLongrangeCorrectedDoublehybrid2009,sancho-garciaAssessmentDoublehybridEnergy2009,zhangDoublyHybridDensity2009,grahamOptimizationBasisSetDependence2009,kozuchDSDBLYPGeneralPurpose2010,goerigkEfficientAccurateDoubleHybridMetaGGA2011,toulouseCommunicationRationaleNew2011,bremondSeekingParameterfreeDoublehybrid2011,sharkasDoublehybridDensityfunctionalTheory2011,chanObtainingGoodPerformance2011,kozuchDSDPBEP86SearchBest2011,zhangFastDoublyHybrid2011,chaiSeekingReliableDoublehybrid2012,mohajeriB2PPW91PromisingDoublehybrid2012,zhangDoublyHybridDensity2012,sharkasMulticonfigurationalHybridDensityfunctional2012}. 
Recently the Møller-Plesset adiabatic connection (MP-AC) approach that recovers MP2 at small coupling strengths and the correct large-coupling strength expansion has been shown to describe the noncovalent interaction well\cite{daasNoncovalentInteractionsModels2021,daasRegularizedOppositeSpinScaled2023}.
To calculate the chemical potential,
the $\Delta$MP2 approach,
which calculates the IP and the EA by the total energy difference at integer electron numbers,
has been used\cite{chongPerturbationCorrectionsKoopmans2003,bacskayCalculationIonisationPotentials1977,besteDirectDMBPTMethod2013,maksicHowGoodKoopmans2002,szaboLinearScalingOpenShellMP22021,wuElectronAffinitiesIonization2006,renResolutionofidentityApproachHartree2012}.
Similar to $\Delta$MP2,
IPs and EAs can be obtained from the total energy difference of double-hybrid functional calculations at integer electron numbers,
which are shown to provide improved accuracy over HF and conventional DFAs\cite{suFractionalChargeBehavior2014}.
MP2 has also been applied in the Green's function formalism.
The second-order Green's function theory (GF2),
which corrects the HF orbital energy by the second-order self-energy,
has also been applied to predict accurate IPs and EAs HF\cite{renRandomphaseApproximationIts2012}.
Recently,
the equation-of-motion MP2 (EOM-MP2) approach has also been developed to calculate IPs and EAs of molecules and solids\cite{langeImprovingMP2Bandgaps2021,langeRelationEquationofMotionCoupledCluster2018}.
However,
IPs and EAs obtained from these approaches are not the rigorous MP2 chemical potential, 
which is the derivative of the MP2 energy with respect to the particle number.

The extension of MP2 to fractional charge and fraction spin systems has been established by Yang et. al\cite{yangExtensionManybodyTheory2013},
which allows to obtain the rigorous MP2 chemical potential.
As shown in Ref.\citenum{yangExtensionManybodyTheory2013},
MP2,
random phase approximation\cite{renRandomphaseApproximationIts2012,eshuisElectronCorrelationMethods2012} (RPA), 
particle-particle random phase approximation\cite{vanaggelenExchangecorrelationEnergyPairing2013,vanaggelenExchangecorrelationEnergyPairing2014} (ppRPA) and a large class many-body perturbation approaches can be expressed as functionals of the non-interacting Green's function.
The fractional formulation of the MP2 correlation energy is achieved by using the ensemble average of the non-interacting Green's function,
which is constructed with occupation-scaled orbitals\cite{yangExtensionManybodyTheory2013}.
In fractional MP2 calculations, 
the fractional charge self-consistent HF calculation is performed first,
then the MP2 correlation energy is evaluated with the HF orbitals obtained for the fractional charge system\cite{cohenFractionalChargePerspective2008}.
With the MP2 correlation energies of the integer and the fractional system,
the MP2 chemical potential can be obtained by the finite difference approach,
which differentiates the MP2 energy expression with respect to the frontier occupation number with the finite difference.
It shows that using the MP2 chemical potential provides better agreements with experiment IP and EA results compared with using HF orbital energy\cite{cohenSecondOrderPerturbationTheory2009}.
By using the fractional formulation of MP2,
the analytical approach to calculate the MP2 chemical potential was developed in Ref~\citenum{suIntegrationApproachSecondOrder2015}.
In the analytical approach,
the derivative of the MP2 correlation energy to the occupation number is evaluated explicitly,
where the orbital relaxation effect is included by solving the coupled-perturbed HF\cite{popleDerivativeStudiesHartreefock1979,frischDirectMP2Gradient1990,suAnalyticDerivativesXYG32013} (CP-HF) equation.
It shows that chemical potentials obtained from the analytical approach agree well with those obtained from the finite difference approach\cite{suIntegrationApproachSecondOrder2015}.
Then the analytical approach was further applied to calculate the chemical potential of double-hybrid functionals,
which shows that the chemical potential of double-hybrid functionals provides smaller errors for predicting IPs and EAs compared with HF and conventional DFAs\cite{suFractionalChargeBehavior2014}. 
The analytical approach only needs the system with an integer electron number and thus avoids systems with fractional charge.
However, 
the analytical approach for MP2 can not be easily extended to other many-body perturbation approaches relying on the non-interacting Green's function.

In the present work,
we introduce a functional derivative approach to calculate the MP2 chemical potential.
As shown in Ref~\citenum{yangExtensionManybodyTheory2013},
the MP2 correlation energy can be expressed as the integration of the non-interacting Green's function and the second-order self-energy on the real frequency axis,
which allows us to calculate the MP2 chemical potential with the chain rule via the non-interacting Green's function.
In the functional derivative approach,
we first take the functional derivative of the MP2 correlation energy with respect to the non-interacting Green's function,
which gives the second-order self-energy.
Then the derivative of the non-interacting Green's function with respect to the occupation number is obtained by solving the CP-HF equation.
We show that the MP2 chemical potentials obtained from the functional derivative approach agree with those obtained from the finite difference approach.
As shown in Section.2 in the Supporting Information,
for MP2 the functional derivative approach is equivalent to the analytical approach in Ref~\citenum{suIntegrationApproachSecondOrder2015}.
However,
the functional derivative approach developed in this work can be easily applied to approximate functionals and many-body perturbation approaches relying on the non-interacting Green's function,
such as RPA and ppRPA.
Previously, 
the self-consistent calculation of perturbation theory based functionals has been performed with the optimized effective potential (OEP) method\cite{mori-sanchezOrbitaldependentCorrelationEnergy2005,smigaSelfconsistentDoublehybridDensityfunctional2016}.
In this work, 
we derive the MP2 Hamiltonian using the functional derivative approach, 
which can be used in the self-consistent calculations of MP2 and double-hybrid functionals in the generalized KS framework, 
with a nonlocal one-electron effective potential.

\section{Methods}
The traditional MP2 correlation energy\cite{szaboModernQuantumChemistry2012,shavittManyBodyMethodsChemistry2009} has been extended to include fractional electrons as\cite{yangExtensionManybodyTheory2013,cohenSecondOrderPerturbationTheory2009}
\begin{equation}\label{eq:mp2_energy}
    E^{\text{MP2}}_{\text{c}} = \frac{1}{4} \sum_{pqrs} n_p n_q (1-n_r) (1-n_s) \frac{\langle pq || rs \rangle \langle rs || pq \rangle}{ \epsilon_p + \epsilon_q - \epsilon_r - \epsilon_s }
\end{equation}
where $n$ is the occupation number, $\epsilon$ is the orbital energy and the two-electron integral is defined as $\langle pq || rs \rangle = \langle pq | rs \rangle - \langle pq | sr \rangle$ with $\langle pq | rs \rangle = \int dx dx' \frac{\psi_p(x) \psi_r(x) \psi_q(x') \psi_s(x')}{|r-r'|}$.
We use $i$, $j$, $k$, $l$ for occupied orbitals, 
$a$, $b$, $c$, $d$ for virtual orbitals and $p$, $q$, $r$, $s$ for general orbitals. 
Eq.\ref{eq:mp2_energy} is initially used as the finite-temperature extension of MP2 with the fractional occupations from finite temperature excitations\cite{ripkaQuantumTheoryFinite1986} and then is derived for fractional systems at zero temperature in Ref.\citenum{yangExtensionManybodyTheory2013}. 

As shown in Ref~\citenum{yangExtensionManybodyTheory2013},
the MP2 correlation energy can be expressed as the integration of the non-interacting Green's function and the second-order self-energy on the real frequency axis
\begin{equation}\label{eq:mp2_energy_green}
    \begin{split}
        E^{\text{MP2}}_{\text{c}} 
        = & \frac{1}{4} \int^{\infty}_{-\infty} \frac{d\omega}{2\pi i} e^{i\omega \eta} 
        \int^{\infty}_{-\infty} \frac{d\omega_1}{2\pi i}
        \int^{\infty}_{-\infty} \frac{d\omega_2}{2\pi i} \\
        & \sum_{pq,rst,uvw} 
        \langle pr || st \rangle \langle uv || qw \rangle
        G^0_{su} (\omega_1) G^0_{tv} (\omega_2) G^0_{wr} (\omega_1 + \omega_2 - \omega) 
        G^0_{qp} (\omega) \\
        = & \frac{1}{4} \int^{\infty}_{-\infty}
        \frac{d\omega}{2\pi i} e^{i\omega \eta} 
        \text{Tr} \{ \Sigma^{(2)} (\omega) G^0 (\omega) \}
    \end{split}
\end{equation}
where $\omega$ is the frequency and $\eta$ is a positive infinitesimal number.
In Eq~\ref{eq:mp2_energy_green} the fractional extension of the non-interaction Green's function $G^0$ in the real space is defined as\cite{yangExtensionManybodyTheory2013}
\begin{equation}\label{eq:green_function}
    G^0(x_1, x_2, \omega) = \sum_i \frac{n_i \psi_i (x_1) \psi^*_i (x_2)}{\omega - \epsilon_i - i\eta}
    + \sum_a \frac{(1-n_a) \psi_a (x_1) \psi^*_a (x_2)}{\omega - \epsilon_a + i\eta}
\end{equation}
Note that the fractional orbital is considered as both the occupied orbital and the virtual orbital.
Thus, the fractional orbitals enters into both occupied and virtual sets in Eq.\ref{eq:green_function}.

The second-order self-energy in Eq~\ref{eq:mp2_energy_green} is defined as\cite{dickhoffManybodyTheoryExposed2008,yangExtensionManybodyTheory2013,renRandomphaseApproximationIts2012}
\begin{equation}\label{eq:2nd_se}
    \Sigma^{(2)}_{pq} (\omega) = \sum_{rst}
    \langle pt || rs \rangle \langle rs || qt \rangle
    \bigg \{
    \frac{(1-n_r)(1-n_s)n_t}{\omega - \epsilon_r - \epsilon_s + \epsilon_t + i\eta} +
    \frac{n_r n_s (1-n_t)}{\omega - \epsilon_r - \epsilon_s + \epsilon_t - i\eta}
    \bigg \}
\end{equation}
which can be separated into two parts:
\begin{equation}\label{eq:sigma_plus}
    \Sigma^{+(2)}_{pq} (\omega) = \sum_{rst}
    \langle pt || rs \rangle \langle rs || qt \rangle
    \frac{(1-n_r)(1-n_s)n_t}{\omega - \epsilon_r - \epsilon_s + \epsilon_t + i\eta}
\end{equation}
and
\begin{equation}\label{eq:sigma_minus}
    \Sigma^{-(2)}_{pq} (\omega) = \sum_{rst}
    \langle pt || rs \rangle \langle rs || qt \rangle
    \frac{n_r n_s (1-n_t)}{\omega - \epsilon_r - \epsilon_s + \epsilon_t - i\eta}
\end{equation}
As shown in Eq~\ref{eq:sigma_plus} and Eq~\ref{eq:sigma_minus},
$\Sigma^{+(2)}$ has poles below the real frequency axis and $\Sigma^{-(2)}$ has poles above the real frequency axis.
The second-order self-energy in Eq~\ref{eq:2nd_se} is used in GF2 to calculate dissociation energies, 
band structures and other properties of molecular and periodic systems\cite{neuhauserStochasticSelfConsistentSecondOrder2017,rusakovSelfconsistentSecondorderGreen2016,phillipsCommunicationDescriptionStrong2014,lanCommunicationInitioSelfenergy2015,renRandomphaseApproximationIts2012,pinoLaplacetransformedDiagonalDyson2004}.

With Eq~\ref{eq:mp2_energy_green}, 
the derivative of the MP2 correlation energy to the occupation number can be obtained from the chain rule via the non-interacting Green's function
\begin{equation}\label{eq:chain_rule}
    \frac{d E^{\text{MP2}}_{\text{c}}}{d n_p} = 
    \int^{\infty}_{-\infty} d \omega 
    \frac{\delta E^{\text{MP2}}_{\text{c}}}{\delta G^0(\omega)} 
    \frac{d G^0(\omega) }{d n_p} 
\end{equation}

As shown in Eq~\ref{eq:mp2_energy_green}, 
the functional derivative of the MP2 correlation energy to the non-interaction Green's function in Eq~\ref{eq:chain_rule} is simply the second-order self-energy
\begin{equation}\label{eq:derivative_e_to_g}
    \frac{\delta E^{\text{MP2}}_{\text{c}} }{\delta G^0_{pq}(\omega)} = \frac{1}{2\pi i} \Sigma^{(2)}_{pq} (\omega) 
\end{equation}

Then with Eq~\ref{eq:green_function}, 
the derivative of the non-interacting Green's function with respect to the occupation number consists of three parts
\begin{equation}\label{eq:derivative_g_to_n}
    \frac{d G^0(\omega) }{ d n_p } =
    \frac{\partial G^0(\omega) }{\partial n_p } +
    \sum_q \frac{\partial G^0(\omega) }{\partial \epsilon_q } \frac{d \epsilon_q}{d n_p} +
    \bigg \{ 
    \sum_q \frac{\partial G^0(\omega) }{\partial \psi_q } \frac{d \psi_q }{ d n_p } 
    + c.c \bigg \}
\end{equation}
where the derivative of the orbital energy to the occupation number
\begin{equation}\label{eq:epsilon_to_n_derivative}
    \frac{d \epsilon_q}{d n_p} = 
    \frac{\partial \epsilon_q}{\partial n_p} + 
    \sum_r \bigg [ 
    \frac{\partial \epsilon_q}{\partial \psi_r} \frac{d \psi_r}{d n_p} + c.c.
    \bigg ]
\end{equation}
and the derivative of the orbital to the occupation number
\begin{equation}
    \frac{d \psi_q}{d n_p} =
    \sum_r \psi_r U^p_{qr}
\end{equation}
are solved from the CP-HF equation\cite{popleDerivativeStudiesHartreefock1979,frischDirectMP2Gradient1990,suAnalyticDerivativesXYG32013}.
In Eq~\ref{eq:epsilon_to_n_derivative} the partial derivative of the orbital energy to the occupation number $\frac{\partial \epsilon_q}{\partial n_p} = \langle qp||qp \rangle $ is also called the ``higher-order term'' in Ref.\citenum{besteDirectDMBPTMethod2013}. 

Then we evaluate three parts in Eq.~\ref{eq:derivative_g_to_n} separately. 
The first part in Eq~\ref{eq:derivative_g_to_n} is the explicit dependence of the non-interacting Green's function on the orbital occupation number
\begin{equation}\label{eq:g_derivative_i}
    \frac{\partial G^0(x_1, x_2, \omega) }{\partial n_p } = 
    -\frac{\psi_p (x_1) \psi^*_p (x_2) }{\omega - \epsilon_p + i\eta}
    + \frac{\psi_p (x_1) \psi^*_p (x_2) }{\omega - \epsilon_p - i\eta}
\end{equation}
The second part in Eq~\ref{eq:derivative_g_to_n} is the dependence of the non-interacting Green's function on the orbital energy
\begin{equation}\label{eq:g_derivative_ii}
    \frac{\partial G^0(x_1, x_2, \omega) }{\partial \epsilon_p } = 
    \frac{(1-n_p) \psi_p (x_1) \psi^*_p (x_2) }{(\omega - \epsilon_p + i\eta)^2}
    + \frac{n_p \psi_p (x_1) \psi^*_p (x_2) }{(\omega - \epsilon_p - i\eta)^2}
\end{equation}
The third part in Eq~\ref{eq:derivative_g_to_n} is the dependence of the non-interacting Green's function on the orbital
\begin{align}\label{eq:g_derivative_iii}
    \frac{\partial G^0(x_1, x_2, \omega) }{\partial \psi_p (x_3) } = & 
    \frac{(1-n_p) \delta(x_1 - x_3) \psi^*_p (x_2) }{\omega - \epsilon_p + i\eta}
    + \frac{n_p \delta(x_1 - x_3) \psi^*_p (x_2) }{\omega - \epsilon_p - i\eta} \\
    \frac{\partial G^0(x_1, x_2, \omega) }{\partial \psi^*_p (x_3) } = &
    \frac{(1-n_q) \psi_p (x_1) \delta(x_2 - x_3) }{\omega - \epsilon_p + i\eta}
    + \frac{n_q \psi_p (x_1) \delta(x_2 - x_3) }{\omega - \epsilon_p - i\eta}
\end{align}

With Eq~\ref{eq:g_derivative_i}, Eq~\ref{eq:g_derivative_ii} and Eq~\ref{eq:g_derivative_iii},
the integral in Eq~\ref{eq:chain_rule} can be performed on the complex plane by using the residue theorem with the contour closing either on the upper half or the lower half plane.
Then the resulting derivative of the MP2 correlation energy with respect to the occupation number consists of the following three parts.

Using Eq~\ref{eq:g_derivative_i}, 
the first part is
\begin{equation}\label{eq:energy_derivative_i}
    \text{I} = \Sigma^{(2)}_{pp} (\epsilon_p)  
    = \Sigma^{+(2)}_{pp} (\epsilon_p) + \Sigma^{-(2)}_{pp} (\epsilon_p)
\end{equation}
which is simply the diagonal element of the second-order self-energy as shown in Ref.\citenum{yangExtensionManybodyTheory2013} and Ref.\citenum{cohenSecondOrderPerturbationTheory2009}. 

Using Eq~\ref{eq:g_derivative_ii}, 
the second part is
\begin{equation}\label{eq:energy_derivative_ii}
    \text{II} = 
    \sum_q \bigg [ 
    n_q \frac{d \Sigma^{+(2)}_{qq} (\omega) }{d \omega} \bigg |_{\omega = \epsilon_q}
    + (1-n_q) \frac{d \Sigma^{-(2)}_{qq} (\omega) }{d \omega} \bigg |_{\omega = \epsilon_q} \bigg ] 
    \frac{d \epsilon_q}{d n_p} 
\end{equation}
where the diagonal element of the first-order derivative of the second-order self-energy to the frequency is
\begin{equation}
    \frac{d \Sigma^{+(2)}_{pp} (\omega) }{d \omega} = 
    - \sum_{rst} \langle pt || rs \rangle \langle rs || pt \rangle
    \frac{(1-n_r)(1-n_s)n_t}{(\omega - \epsilon_r - \epsilon_s + \epsilon_t + i\eta)^2}
\end{equation}
and
\begin{equation}
    \frac{d \Sigma^{-(2)}_{pp} (\omega) }{d \omega} = 
    - \sum_{rst} \langle pt || rs \rangle \langle rs || pt \rangle
    \frac{n_r n_s (1-n_t)}{(\omega - \epsilon_r - \epsilon_s + \epsilon_t - i\eta)^2}
\end{equation}

Using Eq~\ref{eq:g_derivative_iii}, 
the third part is
\begin{equation}\label{eq:energy_derivative_iii}
    \begin{split}
        \text{III} = &
        \sum_q \bigg [ 
        n_q \langle \psi_q | \Sigma^{+(2)} (\epsilon_q) | \frac{d \psi_q}{d n_p } \rangle 
        + (1-n_q) \langle \psi_q | \Sigma^{-(2)} (\epsilon_q) | \frac{d \psi_q}{d n_p } \rangle + c.c \ 
        \bigg ] \\
        = & \sum_q \bigg [ 
        n_q \sum_r \Sigma^{+(2)}_{qr} (\epsilon_q) U^p_{rq} 
        + (1-n_q) \Sigma^{-(2)}_{qr} (\epsilon_q) U^p_{rq} + c.c \ 
        \bigg ] 
    \end{split}
\end{equation}

Combining the above three parts together leads to the full derivative of the MP2 correlation energy to the occupation number
\begin{equation}
    \frac{d E^{\text{MP2}}_{\text{c}}}{d n_p} = \text{I} + \text{II} + \text{III}
    = \Sigma^{+ (2)}_{pp} (\epsilon_p) + \Sigma^{- (2)}_{pp} (\epsilon_p) 
    + \sum_q n_q \frac{d}{d n_p} \Sigma^{+ (2)}_{qq} (\epsilon_q) 
    - \sum_q (1 - n_q) \frac{d}{d n_p} \Sigma^{- (2)}_{qq} (\epsilon_q)
\end{equation}
Because the derivative of the HF total energy to the occupation number is the HF orbital energy\cite{cohenFractionalChargePerspective2008,cohenSecondOrderPerturbationTheory2009},
the MP2 chemical potential, 
which is the derivative of the MP2 total energy to the occupation number,
is given by
\begin{equation}\label{eq:chemical_potential}
    \frac{d E^{\text{MP2}}}{d n_p} = \epsilon^{\text{HF}}_p + \frac{d E_{\text{c}}^{\text{MP2}}}{d n_p} 
\end{equation}
Eq~\ref{eq:chemical_potential} gives the IP when $p$ is the HOMO index and the EA when $p$ is the LUMO index.

The chain rule used for the MP2 correlation energy in Eq.\ref{eq:chain_rule} can be generalized as
\begin{equation}\label{eq:general_chain_rule}
    \frac{d E_{\text{c}}}{d n_p} = 
    \int^{\infty}_{-\infty} d \omega 
    \frac{\delta E_{\text{c}}}{\delta G^0(\omega)} 
    \frac{d G^0(\omega) }{d n_p} 
\end{equation}
Because the derivative of the non-interacting Green's function to the occupation number in Eq~\ref{eq:general_chain_rule} is not dependent on the correlation energy,
it is possible to apply this functional derivative approach to calculate the chemical potential of other approximate functionals and many-body perturbation approaches relying on the non-interacting Green's function. 

Similar to Eq.\ref{eq:chain_rule}, 
the correlation part of the MP2 Hamiltonian can also be derived using the functional derivative approach by taking the derivative of the non-interacting Green's function $G^0$ to the density matrix $\rho$
\begin{equation}
    H^{\text{MP2}}_{\text{c}} = \frac{\delta E^{\text{MP2}}_{\text{c}}}{\delta \rho} = 
    \int^{\infty}_{-\infty} d \omega 
    \frac{\delta E^{\text{MP2}}_{\text{c}}}{\delta G^0(\omega)} 
    \frac{\delta G^0(\omega) }{\delta \rho} 
\end{equation}
As shown in Section.3 in the Supporting Information,
the derivative of the non-interacting Green's function to the density matrix is
\begin{equation}\label{eq:G_to_rho}
    \begin{split}
        & \frac{\delta G (x_1, x_2, \omega)}{ \delta \rho (x_3, x_4) } \\ 
        = & g ( x_1, x_5, \omega + i \eta) f^{\text{Hxc}} (x_5, x_6, x_4, x_3) g ( x_6, x_7, \omega + i \eta) \bar{\rho}_s (x_7, x_2) \\
        & + g ( x_1, x_5, \omega - i \eta) f^{\text{Hxc}} (x_5, x_6, x_4, x_3) g ( x_6, x_7, \omega - i \eta) \rho_s (x_7, x_2) \\
        & + [-g ( x_1, x_5, \omega + i \eta) + g ( x_1, x_5, \omega - i \eta)] \delta (x_3, x_5) \delta (x_2, x_4)
    \end{split}
\end{equation}
where $\rho_s$ is the density matrix,
$\bar{\rho}_s = I - \rho_s$ is the density matrix of the virtual space,
$f^{\text{Hxc}}$ is the Hartree-exchange-correlation (Hxc) kernel is defined as functional derivative of the Hxc potential to the density matrix
\begin{equation}
    f^{\text{Hxc}}_{pq,rs} = \frac{\delta v^{\text{Hxc}}_{pq}}{\delta \rho_{sr}}
\end{equation}
and the function $g (x_1, x_2, \omega )$ is defined as
\begin{equation}
    g (x_1, x_2, \omega) = \sum_p \frac{\psi_p (x_1) \psi_p^* (x_2)}{\omega - \epsilon_p}
\end{equation}
Then the resulting expression of the Hamiltonian is
\begin{equation}\label{eq:mp2_hamiltonian}
    \begin{split}
        [ H^{\text{MP2}}_{\text{c}} ]_{pq} 
        = & \frac{\delta E^{\text{MP2}}_{\text{c}}}{\delta \rho_{pq}}\\
        = & \frac{1}{2} \sum_{ijabr} \frac{ \langle ji || ab \rangle \langle ab || ri \rangle }
        { (\epsilon_a + \epsilon_b - \epsilon_i - \epsilon_r) (\epsilon_a + \epsilon_b - \epsilon_i - \epsilon_j) }
        f^{\text{Hxc}}_{rj,qp} \\
        & + \frac{1}{2} \sum_{ijabr} \frac{ \langle ba || ij \rangle \langle ij || ra \rangle }
        { (\epsilon_i + \epsilon_j - \epsilon_a - \epsilon_r) (\epsilon_i + \epsilon_j - \epsilon_a - \epsilon_b) }
        f^{\text{Hxc}}_{rb,qp} \\
        & + \frac{1}{2} [ \Sigma^{(2)}_{pq} (\epsilon_q) + \Sigma^{(2)}_{pq} (\epsilon_p) ]
    \end{split}
\end{equation}
In Eq.\ref{eq:mp2_hamiltonian},
the last term that contains the second-order self-energy is dominant,
which is similar to the GF2 Hamiltonian with quasiparticle approximation.
Similar to the MP2 correlation energy in Eq.~\ref{eq:mp2_energy} and the second-order self-energy in Eq.~\ref{eq:2nd_se}, 
the MP2 Hamiltonian has a divergence issue for systems with a small or vanished band gap, 
where further studies are needed in future works.

\section{Computational Details}
We implemented the functional derivative approach for MP2 chemical potential in QM4D quantum chemistry package\cite{qm4d}.
In calculations of the correlation part of MP2 chemical potentials obtained from the functional derivative approach and the finite difference approach,
the cc-pVTZ basis set\cite{dunningGaussianBasisSets1989,kendallElectronAffinitiesFirst1992,weigendEfficientUseCorrelation2002} was used for \ce{CH4}, \ce{NH3} and \ce{H2O}.
The cc-pVQZ basis set\cite{dunningGaussianBasisSets1989,kendallElectronAffinitiesFirst1992,weigendEfficientUseCorrelation2002} was used for the remaining atomic systems.
Geometries of \ce{CH4}, \ce{NH3} and \ce{H2O} were taken from the Ref.\citenum{golzeAccurateAbsoluteRelative2020}.
In the finite difference approach,
the difference of the electron number was $10^{-4}$.
In the calculations of IPs and EAs of molecular systems, 
the cc-pVTZ basis set\cite{dunningGaussianBasisSets1989,kendallElectronAffinitiesFirst1992,weigendEfficientUseCorrelation2002} was used.
Geometries and experiment values were taken from Ref.\citenum{vansettenGW100BenchmarkingG0W02015}.
The CCSD(T) results calculated from GAUSSIAN16 A.03 software\cite{g16} were also used as the reference.
All other calculations were performed with QM4D.
QM4D uses Cartesian basis sets and the resolution of identity\cite{eichkornAuxiliaryBasisSets1995,renResolutionofidentityApproachHartree2012,weigendAccurateCoulombfittingBasis2006} (RI) technique to compute two-electron integrals in calculations for the MP2 chemical potential.
All basis sets and corresponding fitting basis sets were taken from the Basis Set Exchange\cite{fellerRoleDatabasesSupport1996,pritchardNewBasisSet2019,schuchardtBasisSetExchange2007}.\\

\section{Results}

\subsection{Validation of the functional derivative approach for the MP2 chemical potential}

\begin{table*}[t!]
\caption{Mean absolute errors (MAEs) of the correlation part of the MP2 correlation energy with respect to the HOMO occupation number obtained from the functional derivative approach at different levels compared with the finite difference approach and the finite difference approach with frozen orbitals.
In the finite difference approach with frozen orbitals,
the MP2 correlation energy of fractional charge systems was evaluated with HF orbitals of integer electron systems.
In finite difference approaches,
the difference of the electron number was $10^{-4}$.
The cc-pVTZ basis set was used for \ce{CH4}, \ce{NH3} and \ce{H2O}.
The cc-pVQZ basis set was used for atomic systems.
Geometries of \ce{CH4}, \ce{NH3} and \ce{H2O} were taken from the Ref.\citenum{golzeAccurateAbsoluteRelative2020}.
All values in eV.}
\label{tab:derivative_homo}\centering
\begingroup
\setlength{\tabcolsep}{10pt}
\begin{tabular}{c|ccccc}
\toprule
          & finite diff & finite diff (frozen) & I     & I+II   & I+II+III  \\
\midrule
\ce{Be}   & -0.27       & -0.56                & -0.56 & -0.47  & -0.27     \\
\ce{B}    & 0.50        & 0.22                 & 0.22  & 0.19   & 0.50      \\
\ce{C}    & 0.83        & 0.60                 & 0.60  & 0.58   & 0.83      \\
\ce{N}    & 1.25        & 1.07                 & 1.08  & 1.05   & 1.25      \\
\ce{O}    & 1.26        & 1.17                 & 1.17  & 1.17   & 1.26      \\
\ce{F}    & 2.11        & 2.12                 & 2.13  & 2.07   & 2.11      \\
\ce{CH4}  & 1.09        & 0.79                 & 0.79  & 0.77   & 1.09      \\
\ce{NH3}  & 1.93        & 1.62                 & 1.63  & 1.56   & 1.93      \\
\ce{H2O}  & 2.73        & 2.57                 & 2.58  & 2.46   & 2.73      \\
\midrule
MAE       &             & 0.27                 & 0.26  & 0.29   & 0.00      \\
\bottomrule
\end{tabular}
\endgroup
\end{table*}

\begin{table*}[t!]
\caption{Mean absolute errors (MAEs) of the correlation part of the MP2 correlation energy with respect to the LUMO occupation number obtained from the functional derivative approach at different levels compared with the finite difference approach and the finite difference approach with frozen orbitals.
In the finite difference approach with frozen orbitals,
the MP2 correlation energy of fractional charge systems was evaluated with HF orbitals of integer electron systems.
In finite difference approaches,
the difference of the electron number was $10^{-4}$.
The cc-pVTZ basis set was used for \ce{CH4}, \ce{NH3} and \ce{H2O}.
The cc-pVQZ basis set was used for atomic systems.
Geometries of \ce{CH4}, \ce{NH3} and \ce{H2O} were taken from the Ref.\citenum{golzeAccurateAbsoluteRelative2020}.
All values in eV.}
\label{tab:derivative_lumo}\centering
\begingroup
\setlength{\tabcolsep}{10pt}
\begin{tabular}{c|ccccc}
\toprule
          & finite diff & finite diff (frozen) & I     & I+II   & I+II+III  \\
\midrule
\ce{Be}   & -0.42       & -0.47                & -0.48 & -0.55  & -0.42     \\
\ce{B}    & -0.96       & -0.96                & -0.96 & -1.07  & -0.96     \\
\ce{C}    & -1.67       & -1.61                & -1.61 & -1.75  & -1.66     \\
\ce{N}    & -2.03       & -2.03                & -2.00 & -2.10  & -2.03     \\
\ce{O}    & -3.19       & -3.19                & -3.05 & -3.22  & -3.19     \\
\ce{F}    & -4.53       & -4.24                & -4.25 & -4.48  & -4.53     \\
\ce{CH4}  & -0.59       & -0.58                & -0.58 & -0.64  & -0.59     \\
\ce{NH3}  & -0.65       & -0.61                & -0.62 & -0.70  & -0.65     \\
\ce{H2O}  & -0.60       & -0.60                & -0.55 & -0.64  & -0.60     \\
\midrule
MAE       &             & 0.07                 & 0.07  & 0.06   & 0.00      \\
\bottomrule
\end{tabular}
\endgroup
\end{table*}

We first examine the correlation part of the MP2 chemical potential obtained from the functional derivative approach at different levels of approximations.
The finite difference approach and the finite difference approach with frozen orbitals were used as the reference. 
In the finite difference approach with frozen orbitals,
the MP2 correlation energy of fractional charge systems was evaluated with HF orbitals of the corresponding integer electron system.
The finite difference of the electron number was set to $10^{-4}$ in two finite difference approaches.
The mean absolute errors (MAEs) of the derivative of the MP2 correlation energy to the HOMO and the LUMO occupation number obtained from the functional derivative approach at different levels of approximations compared with the results obtained from two finite difference approaches are tabulated in Table~\ref{tab:derivative_homo} and Table~\ref{tab:derivative_lumo}.
The first-level approximation (I) only considers the explicit dependence of the MP2 correlation energy on the occupation number,
which is simply the diagonal element of the second-order self-energy as shown in Eq~\ref{eq:energy_derivative_i}.
It shows that the first-level approximation provides an MAE smaller than $0.2$ \,{eV} for the derivative to the HOMO occupation number and smaller than $0.1$ \,{eV} for the derivative to the LUMO occupation number,
which agrees with the results in Ref~\citenum{suIntegrationApproachSecondOrder2015}.
Because the orbital relaxation effect is ignored in the first-level approximation,
the results from the first-level approximation and the finite difference approach with frozen orbitals are very close.
At the first-level approximation,
the functional derivative approach underestimates the derivative to the HOMO occupation number and overestimates the derivative to the LUMO occupation number.
The first-level approximation gives similar results to the orbital energies obtained from diagonalizing the MP2 Hamiltonian defined in Eq.\ref{eq:mp2_hamiltonian}.
Then we examine the second-level approximations (I+II) that further considers the dependence of the MP2 correlation energy on the orbital energy.
It shows that the second-level approximation provides similar or slightly larger MAEs compared with the first-level approximation.
As shown in Table~\ref{tab:derivative_homo},
the second-level approximation further underestimates the derivative to the HOMO occupation number by around $0.05$ \,{eV}.
And in Table~\ref{tab:derivative_lumo}, 
the second-level approximation gives values that are more negative compared with the first-level approximation,
which leads to similar MAEs.
The results of the second-level approximation in this work agree well with the results that includes the dependence on the orbital energy in Ref~\citenum{suIntegrationApproachSecondOrder2015}.
The accurate derivative to the occupation number is obtained at the third-level approximation (I+II+III).
In the third-level approximation, 
the full derivative of the MP2 correlation energy to the occupation number is obtained by further including the dependence on the orbital.
The MAEs of the third-level approximation for calculating the derivative to the HOMO and the LUMO occupation number is $0.0$ \,{eV},
which means the results obtained from the functional derivative approach completely agree with the finite difference approach when the orbital relaxation effect is taken into account.

Therefore,
we demonstrate that the functional derivative approach is capable of predicting accurate chemical potentials of MP2,
which has a simpler form than the analytical approach in Ref~\citenum{suIntegrationApproachSecondOrder2015}.
The equivalence between the functional derivative approach and the analytical approach in Ref~\citenum{suIntegrationApproachSecondOrder2015} for calculating MP2 chemical potentials is shown in Section.2 in the Supporting Information.

\subsection{IPs and EAs obtained from the MP2 chemical potentials}

\begin{table}
    \caption{Mean absolute errors (MAEs) of calculated ionization potentials (IPs) of molecular systems obtained from HF, MP2, $\Delta$HF, $\Delta$MP2 and GF2.
    MP2 stands for the MP2 chemical potential obtained from the functional derivative approach.
    CCSD(T) results obtained from GAUSSIAN16 A.03 software\cite{g16} and experiment values were used as the reference.
    Geometries and experiment values were taken from Ref.\citenum{vansettenGW100BenchmarkingG0W02015}.
    The cc-pVTZ basis set was used.
    All values in eV.}
    \label{tab:ip}\centering
    \begingroup
    \setlength{\tabcolsep}{10pt}
    \begin{tabular}{c|ccccccc}
    \toprule
                   & HF     & $\Delta$HF & MP2    & $\Delta$MP2 & GF2    & CCSD(T) & exp    \\
    \midrule
    \ce{BeO}       & 10.50  & 7.72       & 8.29   & 10.31       & 7.78   & 9.97    & 10.10  \\
    \ce{BN}        & 11.15  & 9.78       & 13.33  & 11.70       & 12.04  & 11.98   &        \\
    \ce{Cl2}       & 12.06  & 11.10      & 10.67  & 11.46       & 11.00  & 11.41   & 11.49  \\
    \ce{CS2}       & 10.13  & 8.73       & 9.28   & 10.67       & 9.83   & 9.99    & 10.09  \\
    \ce{MgF2}      & 15.28  & 13.45      & 11.93  & 14.12       & 11.61  & 13.68   & 13.30  \\
    \ce{F2}        & 18.09  & 15.35      & 13.40  & 17.63       & 13.58  & 15.67   & 15.70  \\
    \ce{Li2}       & 4.95   & 4.35       & 5.02   & 4.93        & 5.21   & 5.22    & 4.73   \\
    \ce{MgCl2}     & 12.23  & 10.65      & 11.10  & 11.89       & 11.31  & 11.64   & 11.80  \\
    \ce{MgO}       & 8.57   & 4.89       & 7.40   & 8.22        & 6.65   & 7.77    & 8.76   \\
    \ce{Na2}       & 4.52   & 4.11       & 4.69   & 4.71        & 4.85   & 4.86    & 4.89   \\
    \ce{NaCl}      & 9.57   & 7.97       & 8.44   & 9.14        & 8.59   & 9.01    & 9.80   \\
    \ce{P2}        & 10.08  & 10.07      & 10.11  & 10.69       & 10.57  & 10.66   & 10.62  \\
    \ce{PN}        & 12.02  & 10.08      & 11.58  & 13.14       & 12.04  & 11.80   & 11.88  \\
    \ce{SO2}       & 13.39  & 11.39      & 10.79  & 13.66       & 11.33  & 12.21   & 12.50  \\
    \midrule
    MAE CCSD(T)    & 0.77   & 1.16       & 0.90   & 0.56        & 0.72   &         &        \\
    MAE exp        & 0.65   & 1.24       & 1.04   & 0.59        & 0.97   & 0.28    &        \\
    \bottomrule
    \end{tabular}
    \endgroup
\end{table}

\begin{table}
    \caption{Mean absolute errors (MAEs) of calculated electron affinities (EAs) of molecular systems obtained from HF, MP2, $\Delta$HF, $\Delta$MP2 and GF2.
    MP2 stands for the MP2 chemical potential obtained from the functional derivative approach.
    CCSD(T) results obtained from GAUSSIAN16 A.03 software\cite{g16} were used as the reference.
    MAEs of all systems and bound systems only are listed separately.
    Geometries were taken from Ref.\citenum{vansettenGW100BenchmarkingG0W02015}.
    The cc-pVTZ basis set was used.
    All values in eV.}
    \label{tab:ea}\centering
    \begingroup
    \setlength{\tabcolsep}{10pt}
    \begin{tabular}{c|cccccc}
    \toprule
                 & HF     & $\Delta$HF & MP2   & $\Delta$MP2 & GF2    & CCSD(T)  \\
    \midrule
    \ce{BeO}     & 1.64   & 2.01       & 1.89  & 1.85         & 2.17   & 1.95    \\
    \ce{BN}      & 2.65   & 4.26       & 5.07  & 2.94         & 3.92   & 2.77    \\
    \ce{Cl2}     & -1.14  & -0.11      & 0.89  & 0.30         & 0.68   & 0.14    \\
    \ce{CS2}     & -1.43  & -0.28      & 0.31  & -0.71        & 0.36   & -0.51   \\
    \ce{MgF2}    & -0.36  & -0.28      & -0.04 & -0.05        & -0.05  & -0.05   \\
    \ce{F2}      & -2.55  & -0.18      & 0.78  & -0.44        & -0.05  & -0.66   \\
    \ce{Li2}     & -0.17  & 0.32       & 0.22  & 0.11         & 0.23   & 0.31    \\
    \ce{MgCl2}   & -0.43  & -0.22      & 0.27  & 0.22         & 0.29   & 0.15    \\
    \ce{MgO}     & 1.23   & 2.79       & 1.78  & -0.08        & 1.25   & 1.36    \\
    \ce{Na2}     & -0.05  & 0.28       & 0.31  & 0.16         & 0.32   & 0.34    \\
    \ce{NaCl}    & 0.47   & 0.53       & 0.57  & 0.57         & 0.59   & 0.55    \\
    \ce{P2}      & -0.65  & 0.00       & 0.53  & -0.05        & 0.42   & 0.02    \\
    \ce{PN}      & -1.33  & -0.35      & -0.14 & -1.41        & -0.28  & -0.65   \\
    \ce{SO2}     & -0.47  & 0.55       & 0.77  & 0.05         & 0.58   & 0.14    \\
    \midrule
    MAE          & 0.60   & 0.38       & 0.55   & 0.26        & 0.36   &         \\
    MAE (bound)  & 0.47   & 0.41       & 0.49   & 0.25        & 0.31   &         \\
    \bottomrule
    \end{tabular}
    \endgroup
\end{table}

Then we examine the performance of using the MP2 chemical potential for predicting IPs and EAs of molecular systems.
The MAEs of calculated IPs and EAs obtained from HF, MP2, $\Delta$HF, $\Delta$MP2 and GF2 compared with CCSD(T) results and experiment results are shown in Table~\ref{tab:ip} and Table~\ref{tab:ea}.
MP2 means the MP2 chemical potential obtained from the functional derivative approach.
In Table~\ref{tab:ea} of the EA results, 
MAEs of all systems and bound systems only are shown separately. 
As shown in the literature\cite{steeleNonCovalentInteractionsDualBasis2009,songDensityCorrectedDFTExplained2022}, 
the prediction of EAs highly depends on basis sets. 
To obtain fully converged EA results, augmented basis sets and extrapolation schemes are needed\cite{steeleNonCovalentInteractionsDualBasis2009,songDensityCorrectedDFTExplained2022}. 
In this work, 
we focus on the comparison between the functional derivative approach and the analytical approach. 
Thus, the cc-pVTZ basis set was used.

For the prediction of IPs,
using HF orbital energies provides a relatively small MAE of $0.77$ \,{eV} for the small molecular systems in the test set.
Because of the lack of correlation effects, 
$\Delta$HF is known to have a poor description for anion systems\cite{suIntegrationApproachSecondOrder2015}.
Thus, $\Delta$HF has a large MAE of $1.16$ \,{eV}.
By including the correlation effects,
$\Delta$MP2 provides the smallest MAE of $0.56$ \,{eV}.
The MAE of GF2 for IPs is slightly larger than $\Delta$MP2,
which agrees with the results in Ref~\citenum{renResolutionofidentityApproachHartree2012}.
The IPs predicted by the MP2 chemical potential provides a relatively large MAE of $0.90$ \,{eV},
because of the deviations of MP2 from the linearity condition\cite{suInsightsDirectMethods2019}.
As shown in Ref~\citenum{suInsightsDirectMethods2019},
the accuracy of using the MP2 chemical potential can be improved by using the two-point formula that averages the derivative to the HOMO occupation number of the $N$-electron system and the derivative to the LUMO occupation number of the $(N-1)$-electron system. 

For the prediction of EAs,
using HF orbital energies provides the largest MAE of $0.60$ \,{eV} for all systems and $0.47$ \,{eV} for bound system.
In addition, 
HF incorrectly predicts most bound systems as unbound systems.
$\Delta$HF provides improvements over HF with smaller MAEs around $0.4$ \,{eV}.
Similar to the prediction of IPs,
$\Delta$MP2 provides the smallest MAE for predicting EAs and correctly describes bound systems except \ce{MgO} and \ce{P2}.
Compared with HF, 
using the MP2 chemical potential correctly describes bound systems.
The EAs obtained from the MP2 chemical potential have larger MAEs compared with $\Delta$MP2,
which is similar to the IP results and can also be improved by using the two-point formula\cite{suInsightsDirectMethods2019}.

\section{Conclusions}

In summary,
we developed a functional derivative approach to calculate the MP2 chemical potential.
By expressing the MP2 correlation energy as an integration of the non-interacting Green's function and the second-order self-energy on the real frequency axis,
the MP2 chemical potential is obtained from the chain rule via the non-interacting Green's function.
First, 
the functional derivative of the MP2 correlation energy with respect to the non-interacting Green's function leads to the second-order self-energy.
Then the derivative of the non-interacting Green's function with respect to the occupation number is obtained by including the orbital relaxation effect.
We showed that the MP2 chemical potential from the functional derivative approach agrees with that from the finite difference approach.
Then the MP2 chemical potential obtained from the functional derivative approach was used to predict IPs and EAs of molecular systems.
It shows that MP2 chemical potentials outperform HF orbital energies for predicting IPs and provides good estimations for EAs.
The MP2 Hamiltonian was also derived using the functional derivative approach,
which can be used in the self-consistent calculations of MP2 and double-hybrid functionals.
The developed functional derivative approach for the MP2 chemical potential can be applied to calculate the chemical potential and the one-electron Hamiltonian of approximate functionals and many-body perturbation approaches relying on the non-interacting Green's function,
which expands the applicability of the Green's function formalism.

\section*{SUPPORTING INFORMATION}
Supporting Information Available:
derivation of the functional derivative approach for the MP2 chemical potential,
equivalence of the analytical approach and the functional derivative approach for the MP2 chemical potential,
derivation of the MP2 Hamiltonian using the functional derivative approach.

\begin{acknowledgement}
J. L. acknowledges the support from the National
Institute of General Medical Sciences of the National Institutes of
Health under award number R01-GM061870. W.Y. acknowledges the support
from the National Science Foundation (grant no. CHE-2154831).
\end{acknowledgement}

\bibliography{ref,software}

\end{document}